\documentclass{article}

\begin{document}

\begin{titlepage}

\begin{center}
\today
\hfill HUB-EP-98/75\\
\hfill hep-th/9812126
\vspace{2cm}

{\Large\bf Equivalence of Geometric Engineering and Hanany-Witten}

\vspace{2cm}
{\large Douglas J.\ Smith}

\vspace{0.5cm}
{\it Humboldt Universit\"at zu Berlin \\
Invalidenstra{\ss}e 110 \\
D-10115 Berlin \\
Germany}

\end{center}

\vspace{1.5cm}

\begin{abstract}
We show the equivalence of three different realisations of gauge theory in
string theory. These are the Hanany-Witten brane constructions, the use of
branes as probes and geometric engineering. We illustrate the equivalence
via T- and S-dualities with the simplest non-trivial examples in four
dimensions: ${\cal N}=2$ SYM with gauge groups $\prod SU(N_i)$.

\end{abstract}

\end{titlepage}

\section{Introduction}
This talk is based on work in collaboration with A.\ Karch and D.\ L\"ust.
More details can be found in \cite{ourpaper}.
Recently there have been three main methods used to describe non-perturbative
supersymmetric gauge theories using
string theory. Each method has its advantages and disadvantages and appears
to be a very different way of describing the same low energy physics.
We will
show that in fact all three methods are equivalent under string dualities. We
begin by describing the Hanany-Witten brane configurations \cite{HW} and
branes at an
orbifold fixed point \cite{probes} which we then relate by T-duality. We
then provide some
evidence for this relation by comparing moduli spaces and deformations at
the classical level. Next we describe the correspondence between
quantum effects in these two descriptions. Finally we relate the branes at
an orbifold to geometric engineering \cite{geom_eng1, geom_eng2, geom_eng3}
via a series of dualities. One of the
crucial ingredients in the duality relations is the use of fractional
branes \cite{FracBranes1, FracBranes2, FracBranes3}.

\section{Hanany-Witten configurations}

In this section we will review the features of the Hanany-Witten \cite{HW}
brane construction for four dimensional ${\cal N} = 2$ gauge theories with
unitary gauge groups \cite{Witten}. We will consider the so-called elliptic
models which means that the 6 direction is compact. Let us first consider
the Higgs branch which is described by $k$ NS5-branes in type IIA string
theory with worldvolume 012345, separated in the 6 direction and at the
origin of the 789 space. We also have $N$ D4-branes wrapped around the 6
direction with worldvolume 01236. When these D4-branes are together but not
at the origin of the 789 space, the low energy theory on their worldvolume
will be ${\cal N} = 4$ SYM with gauge group U($N$).

We can move to the Coulomb branch by taking the D4-branes to the origin of
the 789 space where they can split on the NS5-branes. This gives the
by now familiar ${\cal N} = 2$ Hanany-Witten setup with gauge group U($N$)$^k$
and hypermultiplets in bifundamental representations of each neighbouring
gauge group
\footnote{$k-1$ U(1) factors will be frozen out in the quantum theory but for
simplicity we will mainly discuss the classical theory in this talk.}.
The D4-brane segments which end on consecutive NS5-branes can be
moved independently and so in general we will have gauge group U(1)$^{Nk}$.

\section{T-Duality and Branes on Orbifold}

We can perform at T-duality in the 6 direction on the Higgs branch of the
Hanany-Witten configuration. The D4-branes will become D3-branes with
worldvolume 0123 and the $k$ NS5-branes will become $k$ Kaluza-Klein 5-branes
which we will interpret as a multi-Taub-NUT space \cite{OoguriVafa}. So at
low energies we
have $N$ D3-branes moving in the background of 6789 space with an $A_{k-1}$
singularity at the origin. This can be described by choosing complex
coordinates:
\begin{displaymath}
z_1 = x^6 + ix^7 \hspace{1cm} z_2 = x^8 + ix^9
\end{displaymath}
and modding out by the orbifold action:
\begin{displaymath}
z_1 \rightarrow e^{\frac{2\pi i}{k}}z_1 \hspace{1cm}
z_2 \rightarrow e^{-\frac{2\pi i}{k}}z_2
\end{displaymath}

The low energy theory on the $N$ D3-branes is ${\cal N} = 4$ SYM with gauge
group U(N), exactly as for the T-dual configuration. The D3-branes can be
moved to the orbifold fixed point at the origin of the Higgs branch and then
the low energy theory will be ${\cal N} = 2$ SYM with gauge group U(N)$^k$
which is the T-dual description of the origin of the Higgs branch in the
Hanany-Witten configuration. Now we want to understand how to T-dualise the
Hanany-Witten configuration in the Coulomb branch, where the D4-branes end
on NS5-branes. Our proposal is that the T-dual description will be the
Coulomb branch described by the D3-branes at the orbifold fixed point. The
description of this branch involves the novel concept
of fractional branes \cite{FracBranes1, FracBranes2, FracBranes3}.

Fractional branes are branes with some fraction of the mass and charge of
a full brane. In this example a fractional D3-brane can exist at the
orbifold fixed point with a mass and charge $1/k$ of a full D3-brane. These
fractional branes cannot move away from the fixed point but are free to
move in the 45 directions. $k$ of these fractional branes can combine to
form a complete D3-brane which is then free to move away from the fixed point.
The fractional branes are interpreted as D5-branes wrapping the non-trivial
intersecting 2-cycles of the resolved orbifold fixed point. These 2-cycles
intersect according to the extended
Dynkin diagram for $A_{k-1}$. The sum of these $k$ 2-cycles is a trivial
cycle which explains why $k$ fractional branes can combine to form a
full D3-brane.

It is clear that these fractional branes have the same properties as the
segments of D4-branes ending on NS5-branes. By moving the fractional D3-branes
in the 45 directions we will generically have gauge group U(1)$^{Nk}$. So it
is natural to postulate that this is the T-dual description of the Coulomb
branch of the Hanany-Witten configuration. This can be summarised by the
rule that the T-dual of a D$p$-brane along a direction where it ends on two
consecutive NS5-branes is a fractional D$(p-1)$-brane. Using this rule we
can now
T-dualise a general Hanany-Witten configuration. It should be noted that this
does not require any new assumptions since we can produce a general elliptic
configuration from our special case (with the same number $N$ of D4-branes
between each consecutive NS5-brane) by simply moving some of these D4-brane
segments to infinity in the 45 directions. In this way we can leave any number
$N_i$ of D4-branes between the $i$-th and $(i+1)$-th NS5-branes. This is
exactly reproduced in the T-dual picture by moving the corresponding fractional
D3-branes to infinity in the 45 directions.

\section{Coupling Constants and Fayet-Illiopoulos Terms}

We have already described how the Higgs and Coulomb branches of the
Hanany-Witten setup are mapped to those of the branes in the orbifold space.
Now we will consider the mapping of gauge coupling constants and
Fayet-Illiopoulos terms.

In the Hanany-Witten configuration we can turn on
Fayet-Illiopoulos terms by moving the $i$-th NS5-brane a distance $\zeta_i$
in the 789 space. Since any D4-branes connecting to this NS5-brane will
no longer be parallel to the other D4-branes, this will break supersymmetry.
Indeed it is easy to see that this will increase the energy of the
configuration by an amount of order $\zeta_i^2$ (for small $\zeta_i$). This
means that part of the
Coulomb branch, corresponding to D4-branes ending on the $i$-th NS5-brane,
will be removed. The T-dual description of this is the (partial)
resolution of the orbifold singularity. This is realised by blowing up the
$i$-th 2-cycle and so reducing the singularity type to $A_{k-2}$. This
2-cycle has a radius of order $\zeta_i$ and so a D5-brane wrapping it
will be interpreted as a fractional D3-brane with mass of order $\zeta_i^2$,
the area of the 2-cycle. Clearly this will break supersymmetry and so we again
see that part of the Coulomb branch is removed.

Now we will discuss the T-dual descriptions of the gauge coupling constants.
Consider first the Hanany Witten setup with the $i$-th and $(i+1)$-th
NS5-branes separated in the 6 direction by some distance $\Delta_i$. Then
the gauge coupling of the associated gauge group factor is given by:
\begin{equation}
\frac{1}{g_i^2} = \frac{\Delta_i}{g_s^A l_s}
\end{equation}
where $g_s^A$ and $l_s$ are the type IIA string coupling and string length
respectively. T-dualising this we get the relation between the gauge
coupling and type IIB string coupling:
\begin{equation}
\frac{1}{g_i^2} = \frac{\Delta_i}{g_s^B L_6}
\end{equation}
where $L_6$ is the period around the 6 direction in type IIA.
But we know that for branes in an orbifold background that the gauge coupling
is given by \cite{orb_g}:
\begin{equation}
\frac{1}{g_i^2} \sim \frac{1}{g_s^B}\int_{\sigma_i}{\cal F}
\end{equation}
where $\sigma_i$ is the $i$-th 2-cycle and ${\cal F} = F - B^{NS}$ is the
gauge invariant 2-form field strength on the D5-brane (as appears in the
Born-Infeld action.) So we see the expected relations from T-duality: the
6 position of an NS5-branes is translated into a Wilson line. So the
seemingly different determinations of the gauge coupling are actually just
T-dual.

\section{Comments on Quantum Effects}

Although what we have described so far has been at the classical level, the
T-duality should also relate quantum properties of the two descriptions of
the gauge theory. For ${\cal N} = 2$ theories in four dimensions there are
two types of quantum effects: 1-loop effects and non-perturbative instanton
contributions.

In the Hanany-Witten description the gauge theory instantons are Euclidean
D0-branes whose worldline is in the 6 direction \cite{Brodie}. They end on
the NS5-branes
and so they should T-dualise into fractional D($-1$)-branes at the orbifold
fixed point. This is indeed what is expected since D-instantons on D3-branes
are simply Yang-Mills instantons.

The 1-loop effects have a simple geometric interpretation in the Hanany-Witten
setup. Here the NS5-branes bend logarithmically since the D4-branes ending on
them have a tension. The logarithmic behaviour is simply because the ends of
the D4-branes have codimension 2 on the NS5-branes. Since the distance between
the NS5-branes determines the gauge coupling the variation of the distance
is interpreted as the running of the gauge coupling. In the T-dual picture
the distance between the NS5-branes becomes a Wilson line for ${\cal F}$ which
again determines the gauge coupling. The logarithmic running can be seen
directly since the fractional D3-branes are charged objects in the
2-dimensional space transverse to them and the orbifold. This means that the
4-form potential has a logarithmic dependence. Since this couples to
${\cal F}$ in the D5-brane action, when we compactify the D5-brane on a
2-cycle to get the fractional D3-brane we will find that the effective gauge
coupling has a logarithmic dependence.

\section{From Branes on Orbifold to Geometric Engineering}
It is now quite straightforward to complete the duality relation between the
Hanany-Witten description and geometric engineering. If we consider the
situation with an $A_{k-1}$ singularity and $N_i$ fractional D3-branes,
i.e. $N_i$ D5-branes wrapped on each 2-cycle $\sigma_i$, then we can make
an S-duality transformation. This doesn't affect the singularity but will
transform the D5 branes into NS5-branes wrapping the same cycles. We can now
perform a T-duality in one of the two directions transverse to the NS5-branes
and the orbifold. This will transform the NS5-branes into $A_{N_i-1}$
singularities fibred over the 2-cycles of the original $A_{k-1}$ singularity.
This is precisely the description of a non-compact Calabi-Yau threefold
used in geometric engineering to describe the same gauge theory we have been
considering. More details of this duality can be found in \cite{ourpaper}.

\section{Conclusions}
We have shown the dualities connecting Hanany-Witten setups, via branes at an
orbifold, to geometric engineering in this simplest non-trivial example. This
can easily be extended to more general examples of gauge theories in various
dimensions and with orthogonal or symplectic groups by including
orientifold planes. Theories with reduced supersymmetry can also be analysed
in this way. For example the brane box models \cite{BraneBox} have been
studied by dualising
to branes at orbifolds using this method \cite{BoxOrb}. There are of course
some limitations
to each method. For example exceptional groups cannot be constructed using
the Hanany-Witten setup but can by using geometric engineering. This is
translated here into the fact that we cannot T-dualise an E-type singularity.


\begin{thebibliography}{3}
\bibitem{ourpaper}
A.\ Karch, D.\ L\"ust, D.J.\ Smith
{\it Equivalence of geometric engineering and Hanany-Witten via
fractional branes}
Nucl.\,Phys. {\bf B533} (1998) 348-372; hep-th/9803232
\bibitem{HW}
A.\ Hanany, E.\ Witten
{\it Type IIB Superstrings, BPS monopoles, and three-dimensional gauge dynamics}
Nucl.\,Phys. {\bf B492} (1997) 152-190; hep-th/9611230
\bibitem{probes}
M.R.\ Douglas, G.\ Moore
{\it D-branes, quivers, and ALE instantons}
hep-th/9603167
\bibitem{geom_eng1}
A.\ Klemm, W.\ Lerche, P.\ Mayr, C.\ Vafa, N.\ Warner
{\it Selfdual strings and N=2 supersymmetric field theory}
Nucl.\,Phys. {\bf B477} (1996) 746-766; hep-th/9604034
\bibitem{geom_eng2}
S.\ Katz, A.\ Klemm, C.\ Vafa
{\it Geometric engineering of quantum field theories}
Nucl.\,Phys. {\bf B497} (1997) 173-195; hep-th/9609239
\bibitem{geom_eng3}
S.\ Katz, P.\ Mayr, C.\ Vafa
{\it Mirror symmetry and exact solution of 4-D N=2 gauge theories -- I}
Adv.\,Theor.\,Math.\,Phys. 1 (1998) 53-114; hep-th/9706110
\bibitem{FracBranes1}
J.\ Polchinski
{\it Tensors from K3 orientifolds}
Phys.\,Rev. D55 (1997) 6423-6428; hep-th/9606165
\bibitem{FracBranes2}
M.R.\ Douglas
{\it Enhanced gauge symmetry in M(atrix) theory}
J.\,High Energy Phys. 07 (1997) 004; hep-th/9612126
\bibitem{FracBranes3}
D.-E.\ Diaconescu, M.R.\ Douglas, J.\ Gomis
{\it Fractional branes and wrapped branes}
J.\,High Energy Phys. 02 (1998) 013; hep-th/9712230
\bibitem{Witten}
E.\ Witten
{\it Solutions of four-dimensional field theories via M theory}
Nucl.\,Phys. {\bf B500} (1997) 3-42; hep-th/9703166
\bibitem{OoguriVafa}
H.\ Ooguri, C.\ Vafa
{\it Two-dimensional black hole and singularities of CY manifolds}
Nucl.\,Phys. {\bf B463} (1996) 55-72; hep-th/9511164
\bibitem{orb_g}
A.\ Lawrence, N.\ Nekrasov, C.\ Vafa
{\it On conformal field theories in four-dimensions}
Nucl.\,Phys. {\bf B533} (1998) 199-209; hep-th/9803015
\bibitem{Brodie}
J.\ Brodie
{\it Fractional branes, confinement, and dynamically generated superpotentials}
Nucl.\,Phys. {\bf B532} (1998) 137-152; hep-th/9803140
\bibitem{BraneBox}
A.\ Hanany, A.\ Zaffaroni
{\it On the realization of chiral four-dimensional gauge theories using branes}
J.\,High Energy Phys. 05 (1998) 001; hep-th/9801134
\bibitem{BoxOrb}
A.\ Hanany, A.M.\ Uranga
{\it Brane boxes and branes on singularities}
J.\,High Energy Phys. 05 (1998) 013; hep-th/9805139

\end{thebibliography}
\end{document}